\documentclass[12pt]{elsart}

\usepackage{amssymb}

\begin{document}

\begin{frontmatter}


\begin{center}
{\scriptsize Physica A 324 (2003) 723--732}
\end{center}

\title{Intrinsic chaos and external noise in population dynamics}
\author[jo1,jo2]{Jorge A. Gonz\'alez},
\author[jo1,le,le1]{Leonardo Trujillo},
\ead{leo@pmmh.espci.fr}
\author[an]{Anan\'{\i}as Escalante}
\address[jo1]{The Abdus Salam International Centre for Theoretical Physics (ICTP),
Strada Costiera 11, 34100, Trieste, Italy}
\address[jo2]{Centro de F\'{\i}sica, Instituto Venezolano de Investigaciones
Cient\'{\i}ficas, A.P. 21827, Caracas 1020-A, Venezuela.}
\address[le]{P. M. M. H. Ecole Sup\'erieure de Physique et Chimie Industrielles, 10 rue Vauquelin,
75231 Paris Cedex 05, France.}
\corauth[le1]{Corresponding author. Fax: +33--1--40--79--45--23}
\address[an]{Laboratorio de Ecolog\'{\i}a  de Poblaciones, Centro de Ecolog\'{\i}a, Instituto Venezolano de Investigaciones
Cient\'{\i}ficas, A.P. 21827, Caracas 1020-A, Venezuela.}

\begin{abstract}
We address the problem of the relative importance of the intrinsic chaos
and the external noise in determining the complexity of population dynamics.
We use a recently proposed  method for studying the complexity of
nonlinear random dynamical systems. The new measure of complexity is defined
in terms of the average number of bits per time--unit necessary to
specify the sequence generated by the system. This measure coincides with
the rate of divergence of nearby trajectories under two different realizations
of the noise. In particular, we show that the complexity of a nonlinear
time--series model constructed from sheep populations comes completely
from the environmental variations. However, in other situations, intrinsic
chaos can be the crucial factor. This method can be applied to many other
systems in biology and physics.
\end{abstract}

\date{{\footnotesize  }}
\end{frontmatter}


\section{Introduction}

Recently several outstanding
papers\cite{Castro,Dimitrova,Droz,Penna,Sznajd,Sole,Monetti,Johnson} have
applied physical and mathematical methods to ecology and population dynamics.
This is a very important development. In fact, interdisciplinary research can
produce very significant ideas.

There exists a great controversy in ecology\cite{May1,May2,Sugihara1,Ellner,Sugihara2,Grenfell1,Stenseth}
concerning the relative
importance of intrinsic factors and external environmental variations in
determining populations fluctuations. In this article we address this problem using a
recently proposed method\cite{Paladin,Loreto} for studying the complexity of  a nonlinear
random dynamical system. This method characterizes the complexity by
considering the rate $K$ of divergence of nearby orbits evolving under two
different noise realizations. We can show that this measure is very effective
for investigating nonlinear random systems. In Ref.\cite{Grenfell1} a nonlinear
time-series model is constructed from sheep populations on two islands in
the St. Kilda archipelago\cite{Ranta,Grenfell2}. We investigate the complexity of this
model using the new technique. We have shown that the complexity of the system
comes completely from the environmental variations. This combination of new
methods is a very powerful tool for quantifying the impact of environmental
variations on population dynamics and can be applied to other systems.

The paper is organized as follows: In Section 2 we recall the definition of
complexity for random dynamical systems. In Section 3 we recall the definition
and properties of the random sequences given in
Ref.\cite{Jorge1,Jorge2,Jorge3,Jorge4}, and we compute the complexity for the random
sequences and a particular random map. In Section 4 we discuss a nonlinear
time--series model constructed from sheep population data and we compute the
complexity for this model. In this section we also show that
for a generalized model, the complexity can depend on both, the intrinsic chaos
and the environmental variations. In Section 5 we briefly discuss some aspects of
the problem of distinction between deterministic chaos and noise.

\section{Complexity in random dynamical systems}

Recently a new measure of complexity was introduced\cite{Paladin,Loreto} in
terms of the average number of bits per time-unit necessary to specify the
sequence generated by the system. This definition becomes crucial in random
nonlinear dynamical systems as the following
\begin{equation}
X_{n+1}=f(X_n,I_n),
\label{eq:1}
\end{equation}
where $I_n$ is a random variable (e.g. noise).  This measure coincides
with the rate $K$ of divergence of nearby trajectories under two different
realizations of the noise. The method of calculating the Kolmogorov-Sinai
entropy with the separation of two nearby trajectories with the same
realization of the noise can lead to incorrect results\cite{Paladin,Loreto}. The
complexity of the dynamics generated by (\ref{eq:1}) can be calculated as
\begin{equation}
K=\lambda\theta (\lambda)+h,
\label{eq:2}
\end{equation}
where $\lambda$ is the Lyapunov exponent of the map, which is defined as
\begin{equation}
\lambda = \lim_{n\rightarrow \infty} n^{-1}\ln |Z_n|,
\end{equation}
where $Z_{n+1}= (\partial f(X_n)/\partial X_n)Z_n$,
$h$ is the complexity
of $I_n$,
(which can be calculated as the Shannon entropy of the sequence $I_n$),
and $\theta (\lambda)$ is the Heaviside step function. This
function is defined as follows: $\theta (\lambda) = 0$  if $\lambda\leq 0$;
$\theta (\lambda) = 1$ if $\lambda > 0$.
For a detailed explanation of the relationship between the definition of
complexity as the average number of bits per time unit necessary to
specify the sequence, the rate of divergence of nearby trajectories under
two different realizations of the noise and the equation (\ref{eq:2}), see
Ref.\cite{Boffetta}. The definition in this form was given in the original
papers\cite{Paladin,Loreto}. However, in a different formalism Eq.(\ref{eq:2}) could be
considered as the starting definition of $K$.
On the other hand, there are many alternative
measures of complexity. So we should check the effectiveness of this new
method. In the next section, using some random sequences and a random map,
we will show that the rate of divergence of nearby trajectories under
two different realizations of the noise  indeed can be calculated using
equation (\ref{eq:2}).

\section{Random sequences}

Very recently\cite{Jorge1,Jorge2,Jorge3,Jorge4} we have investigated
explicit functions which can produce truly random numbers
\begin{equation}
X_n=\sin ^2(\theta\pi Z^n). \label{eq:3}
\end{equation}
When $Z$ is an integer, function (\ref{eq:3}) is the exact solution to chaotic maps.
However, when $Z$ is a generic fractionary number, this is a random function
whose values are completely independent. Using these functions (or an
orthogonal set of them) we can find exact solutions to random maps as
equation (\ref{eq:1}).

Let us discuss some properties of function (\ref{eq:3}). Let $Z$ be a
rational number expressed as $Z=p/q$, where $p$ and $q$ are relative prime
numbers. We are going to show that if we have $m+1$ numbers generated by
function (\ref{eq:3}): $X_{0}, X_{1}, X_{2}, X_{3}, ...,X_{m}$ ($m$ can be
as large as we wish), then the next value $X_{m+1}$, is still unpredictable.
This is valid for any string of $m+1$ numbers. Let us define the following
family of sequences
\begin{equation}
X_{n}^{(k,m)}= \sin^2 \left[  \pi \left ( \theta_0 + kq^{m} \right ) \left( p/q \right )^n \right ],
\label{eq:4}
\end{equation}
where $k$ and $m$ are integer. For all sequences parametrized by $k$, the
first $m+1$ values are the same. This is so because
\( X_{n}^{(k,m)}= \sin^2 \left[  \pi\theta_0 ( p/q )^n + \pi k p^n q^{(m-n)}  \right ] = \sin^2\left[ \pi\theta_0 (p/q)^n \right] \),
for all $n\le m$.
Nevertheless, the next value
\begin{equation}
X_{m+1}^{(k,m)}=\sin^2 \left[ \pi\theta_0 (p/q)^{m+1} + \frac{\pi k p^{m+1}}{q}\right],
\label{eq:5}
\end{equation}
is unpredictable. In general, $X_{m+1}^{(k,m)}$ can take $q$ different
values. These $q$ values can be as different as $0, 1/5, 1/2, \sqrt{2}/2$ or
$1$. For $Z$ irrational there can be an infinite number of different
outcomes.

Function (\ref{eq:3}) with $Z=3/2$ (i.e. $X_n = \sin^2 \left[\theta
\pi(3/2)^n\right]$) is a solution of the following map
\begin{equation}
X_{n+1}= \frac{1}{2}\left[ 1 + I_n \left( 1 - 4X_n \right)\left( 1-X_n
\right)^{1/2} \right], \label{eq:6}
\end{equation}
where
\begin{equation}
I_n = -\frac{\cos\left[ \theta \pi \left( 3/2 \right)^n \right]}{\left\{ 1 - \sin^2 \left[ \theta\pi \left( 3/2 \right)^n \right] \right\}^{1/2}}
\label{eq:7}
\end{equation}
if \( \sin^2 \left[ \theta\pi \left( 3/2 \right)^n \right] \neq 1\); and
\begin{equation}
I_n = 1,
\label{eq:8}
\end{equation}
if  \(\sin^2 \left[ \theta\pi \left( 3/2 \right)^n \right] =1\).

A careful analysis of function $I_n$ yields that $I_n$ is an
unpredictable function that takes the values $\pm 1$ with equal
probability. A particular realization of $I_n$ for $\theta = 0.77$
is the following: 1, -1, 1, -1, -1, 1, 1, -1, -1, -1, 1, -1, -1,
1, 1, -1, 1, -1, 1, ... The same analysis of Eq.(\ref{eq:4}) made
above (in this case for $Z=3/2$) confirms these results. On the
other hand, a statistical investigation of the outcomes of
function $I_n$ corroborates these findings. In fact, it does not
matter how many past values $I_0, I_1, I_2..., I_m$ we already
know, the next value cannot be determined. It can be either 1 or
-1 with the same probability. In other words, $I_n$ behaves as a
random coin toss.

Now we can check some of the results discussed by the authors of
Ref.\cite{Paladin,Loreto}. In the case of the random map
(\ref{eq:6}) $\lambda = \ln (3/2)$ and $h=\ln 2$. Thus, $K=\ln 3$.
Here we give a brief explanation of these results. The map
(\ref{eq:6}) can be re--written on the following form
\begin{eqnarray}
X_{n+1}=\left\{ \begin{array}{ll}
\frac{1}{2}\left[ 1 + (1 - 4X_n)(1 - X_n)^{1/2}\right],  &  \qquad \mbox{with probability $\frac{1}{2}$}, \\
\frac{1}{2}\left[ 1 - (1 - 4X_n)(1 - X_n)^{1/2}\right],  &  \qquad \mbox{with probability $\frac{1}{2}$}.
\end{array}
\right.
\end{eqnarray}
This is a form also compatible with the application of equation (2) (See
\cite{Paladin,Loreto}).  After the transformation  $X_n = \sin^2(Y_n) $, both
equations \begin{equation}
X_{n+1} = \frac{1}{2}\left[ 1 + (1 - 4X_n)(1 - X_n)^{1/2}\right],
\end{equation}
and
\begin{equation}
X_{n+1} = \frac{1}{2}\left[ 1 - (1 - 4X_n)(1 - X_n)^{1/2}\right],
\end{equation}
can be converted into piece--wise linear maps where the absolute value of
the slope $\left| dY_{n+1}/dY_n \right|$ is constant and equal to $3/2$.

Using the equation (3) for the Lyapunov exponent, we obtain  the exact value
$\lambda = \ln(3/2)$. The Lyapunov exponent is invariant with respect to the
transformation $X_n = \sin^2(Y_n)$. If we calculate numerically the
Lyapunov exponent of maps (7), (11) and (12), we also obtain the value
$\lambda = \ln(3/2)$ approximately.

Considering the properties of the sequence $I_n$ (that takes the values $1$ and $-1$
with equal probability), it is trivial to get that $h= \ln 2$. Applying equation (2),
we obtain $K = \ln 3$. All these calculations have been made
using the definitions of the quantities and the algebraic structure of
equation (7). Now let us consider the analytical solution of map (7):
\begin{equation}
X_n = \sin^2\left[ \theta \pi (3/2)^n \right].
\end{equation}
If we investigate equation (13), it is possible to prove that, on average, for
a given $\theta$, nearby trajectories will separate following the law
$d \sim (3/2)^n$, where $d$ is the distance between the trajectories. This
yields that $\lambda = \ln(3/2)$, which corroborates a previous result.

A more important calculation is that of $K$. We wish to compute the
rate of divergence of nearby trajectories under two different
realization of the noise. In the ``language'' of the exact solution (13)
this is equivalent to investigate the average divergence of trajectories that
are very close for $n=0$, but with different values of $\theta$ (recall that
different realizations of the random variable $I_n$ (equation (8)) are
produced with different values of $\theta$. This analysis yields the
following result $K = \ln 3$. And this is a corroboration of equation (2) for this
system.

Now let us resort to numerical calculations. We have produced numerically
$10000$ values of $X_n$ using both, the dynamical system (7) and the
function (13). Then, we have computed the complexity
of these sequences using the Wolf's algorithm\cite{Wolf}. The result
is very close to $\ln 3$ (in fact $K \approx 1.098$).
Moreover, even an independent calculation of the complexity of this
dynamics using different methods\cite{Wolf,Pincus} produces the same result.
In Ref.\cite{Pincus} a new method for the calculation of the complexity
of a sequence is developed. This method has been shown to be very
effective for the calculation of complexity of finite sequences\cite{Pincus,Pincus2,Singer,Pincus3}.
We start with a sequence of values $U_1,U_2,U_3,...,U_N$; from  which we can
form a sequence of vectors
\begin{equation}
X(i) = \left[U_i,U_{i+1},...,U_{i+m-1} \right].
\end{equation}
Now, we will define some variables:
\begin{equation}
C_i^m(r) = \frac{ (\mbox{number of $j$ such that } d\left[ X(i),X(j) \right]\leq r)}{(N-m+1)},
\end{equation}
where $d[X(i),X(j)]$ is the distance between two vectors, which is defined as follows:
\begin{equation}
d[X(i),X(j)]= \mbox{max} (|U_{i+k-1}-U_{j+k-1}|), \qquad (k=1,2,..,m).
\end{equation}
Another important quantity is
\begin{equation}
\phi^m(r) = \sum_{i=1}^{N-m+1} \frac{ \ln C_i^m(r)}{N-m+1}.
\end{equation}
Using all these definitions, we can calculate the complexity
\begin{equation}
K(m,r,N)= \phi^m(r)-\phi^{m+1}(r).
\end{equation}
This measure depends on the resolution parameter $r$ and the
{\it embedding} parameter m, and represents a computable
framework for the ``Shannon's entropy'' of a finite real sequence.

It is interesting that when we calculate numerically the complexity of the
sequences produced by the map (7) and the exact solution (13) (with $r=0.025$,
and different $m\geq 2$) we obtain $K\approx 1.098$.

Using functions (\ref{eq:3}) (with different values of $Z$)
we can also solve maps where the Lyapunov
exponent is negative and, nevertheless, due to the existence of
external noise, the complexity is positive. In the presence of random
perturbations, $K$ can be very different from the standard Lyapunov
exponent and, hence, from the Kolmogorov--Sinai entropy computed
with the same realization of the noise.

In general, if we apply the measure of complexity $K$ to our
functions (\ref{eq:3}), then we obtain the following results: for
$Z=p/q$, $K=\ln p$. If $Z$ is irrational, the complexity is
infinite. We should add some comments about these computations.
When $Z$ is integer, function (4) is equivalent to a univalued
chaotic map of type $X_{n+1} = f(X_n)$. In this case, this measure
coincides with the Kolmogorov--Sinai entropy i.e. $K = \lambda$.
So $\lambda = \ln Z$. When $Z = p/q$, where $p$ and $q$ are
relative primes, function (4) produces multivalued firsts-return
maps\cite{Jorge1,Jorge2,Jorge3,Jorge4}. In this case, the
information lacking is not given by $K = \lambda$, but is larger:
one loses information not only in each iteration due to
$\lambda>0$. One has also to specify the branch of the map
$(X_n,X_{n+1})$ in each iteration. We should apply the formula
(2), where $h$ is the entropy of the random jumps between the
different branches of the map $(X_n,X_{n+1})$. For $Z= p/q$, there
are q branches in the map $(X_n,X_{n+1})$. Investigating the
properties of function (4) we arrive at the  conclusion that all
the branches possess the same probability in the process of
jumping. This leads to the equality $h = \ln q$. Thus, $K = \ln
(p/q) + \ln q = \ln p$. The complexity can be obtained by
computing the separation rate of nearby trajectories evolving in
two different realizations of the noise. In the case of sequences
produced by function (14), this is equivalent to using two
different $\theta$ for which (at $n = 0$) the trajectories are
close. Such procedure exactly corresponds to what happens when
experimental data are analyzed with the Wolf {\it et al.}
algorithm \cite{Wolf}. When we apply the Wolf {\it et al.}
algorithm to the sequences generated by our functions and the
mentioned dynamical systems, we obtain the expected theoretical
results. The complexity of the sequences produced by function (4)
also can be calculated using random dynamical systems for which
function (4) is the exact solution as we did in the case $Z=3/2$.
For different values of $Z$, the result is again $K= \ln p$.

\section{Sheep population model}

In a beautiful work Grenfell {\it et al.}\cite{Grenfell1} used the unusual
situation of time series from two sheep populations that were very close (and
so, they shared approximately the same environmental variation as for example
rain, temperature, etc) but which were isolated from each other, i.e. these
populations did not interact, in order to study the interaction between noise
and
nonlinear population dynamics.
They found high correlations in the two sheep
populations on two islands in the St. Kilda archipelago. They were able to
express $X_{n+1}$ as a function of the previous population size:
$X_{n+1} = f(X_n) + \epsilon_{n+1}$, where $\epsilon_n$ represents
the noise, which is related to the environmental variables
($n$ is the discrete time).
Here $X_n = \log N_n$, where $N_n$ is the population number. They fit a
nonlinear self--exciting threshold autoregressive (SETAR) model\cite{Tong1,Tong2,Tsay}
to the Hirta island (one of the islands of the St. Kilda archipelago) time
series. The best-fit model is
\begin{eqnarray}
\begin{array}{ll}
X_{n+1}=a_0 + b_0 X_n +\epsilon_{n+1}^{(0)}, & X_n \leq c, \\
X_{n+1}=a_1 +\epsilon_{n+1}^{(1)},           & X_n > c,
\end{array}
\label{eq:9}
\end{eqnarray}
where $c = 7.066$; $a_0 = 0.848$; $b_0 = 0.912$; $\sigma_0 = 0.183$;
$a_1 = 7.01$; $\sigma_1 = 0.293$. Here $\sigma_{0,1}$ is the variance of
$\epsilon_n$.
The noise $\epsilon_n$ is defined as a sequence of independent  and
identically distributed normal random numbers with mean $0$ and
variance $\sigma$.
The model captures the essential features of the time--series,
including the map $X_{n+1}$ versus $X_n$. Now we apply the measure of
complexity $K$ (Eq.(\ref{eq:2})) to the model given by equation (\ref{eq:9}). It is
straightforward to show that $\lambda < 0$. This can be done even
analytically using equation (3).
Thus, the complexity of the dynamical system (\ref{eq:9}) is $K=h$, where $h$ is the
complexity of the noise $\epsilon_n$. That is, all the complexity of this
dynamical system comes completely from the environmental variations. We
could say that, in this case, the extrinsic environmental variations are
much more important than the intrinsic factors in determining population
size fluctuations.

This result confirms the results of Ref.\cite{Grenfell1}. However, we should
note that their research is based on the very particular situation
where we have synchronization of two population fluctuations at separate,
but not too distant locations.

Here we should explain briefly how Grenfell {\it et al.}\cite{Grenfell1}
obtained their results. They found that the fluctuations in the sizes of the
two populations are remarkably synchronized over a 40-year period. They
explain this synchronization using the fact that the two populations
are exposed simultaneously to the same environmental variations. Assuming
that the same model applies to both islands, they use it to estimate the
level of correlation in environmental
noise required to
generate the observed synchrony in population fluctuations. They found that
very high levels of noise correlation are needed to generate the observed
correlation between the sheep populations on the two islands. They also
studied observed large--scale meteorological  covariates like monthly
wind, rain and temperature. From this analysis they conclude that the
extrinsic influences are very important in this particular case of population
dynamics.

On the other hand, our method can be
applied to any other population dynamics.
Even if we have only one isolated population in the same region.
The research program is the
following: the data should be fitted by a SETAR model and, after that,
the complexity can be calculated using equation (\ref{eq:2}). In fact, many nonlinear
population models can be approximated by a SETAR model. This is a very
clear theoretical result.  It does not depend on further statistical
assumptions or approximate investigations. Once we have reconstructed the
model from the data (and this is a step that we cannot avoid in any other
method), we can prove rigorously that the Lyapunov exponent is negative
and that all the complexity comes from the external random perturbations.
Our results explain why the
environmental variations are more important in this particular case.
This is due to the density--dependent relationship  $X_{n+1} = f(X_n)$.
In fact, for other animal populations the best--fit model can be very
different. The form of the density dependence is crucial. For instance,
suppose that the best--fit model is similar to that presented in Ref.\cite{Stenseth}:
\begin{eqnarray}
X_{n+1}=\left\{ \begin{array}{ll}
r+ X_n + \epsilon_{n+1}^{(1)}, & X_n \leq c, \\
(r + b c) + (1-b)X_n + \epsilon_{n+1}^{(2)}, & X_n > c.
\end{array}
\right.
\end{eqnarray}

Here we should add a short explanation of the origin of Eq.(11). In
Ref.\cite{Maynard} Mayrand Smith and Slatkin discuss the so--called MSS model
\begin{equation}
N_{n+1} = \frac{RN_n}{(1+N_n/N_c)^b},
\end{equation}
where $N_n$ is the population size at time $n$, $R$ is the maximal net population
growth rate, $b$ is a measure of the strength of the density dependent reduction
of the net population growth rate, and $N_c$ is the carrying capacity (that is,
the maximum population size that can be sustained by the area under study).
If we introduce the transformation $X_n = \ln N_n$ (and the effect of noise), we
can obtain equation (20) as an approximation, where $r=\ln R$, $c=\ln N_c$
(See Ref.\cite{Stenseth}). However, in the same way as Eq.(19), Eq.(20) can be
obtained as a SETAR model reconstruction using a given time--series
of the evolution of certain animal population. In general, as pointed
out by Stenseth and Chan\cite{Stenseth}, many nonlinear population models may, on the
$\log$--scale, be approximated by a dynamical system similar to
equation (20).

For large values of parameter $b$, the Lyapunov exponent can be positive.
In  this case, both the intrinsic and the external factors contribute to the
variability of the dynamics.  Moreover, it can be that the intrinsic chaotic
factors are the most important in determining population fluctuations.
Nevertheless, we have shown that randomness is crucial in ecological models.

\section{Chaos and noise}

In the research presented in this paper the question of chaos or noise is very
relevant both in the problems related to the random functions and in the
problems of characterizing experimental time-series.

We should say here that recently several important papers have been dedicated
to the question of distinguishing between generic deterministic chaos and
noise\cite{Sugihara,Wales,Tsonis,Boffetta,Cencini}.
Several limitations have been found for the usual methods that
are based on the calculation of the Lyapunov exponent and the Kolmogorov--Sinai
entropy\cite{Cencini}. Many of the practical problems are related to the fact that
these quantities are defined as infinite time averages taken in the limit
of arbitrary fine resolution.

New very strong methods have been developed based on different concepts. Some
of these methods\cite{Sugihara,Wales,Tsonis} are based on the differences
in the predictability when
time--series is analyzed using prediction algorithms.

In Ref.\cite{Cencini} this problem is solved by introducing the $(\epsilon ,\tau )$
entropy ($h(\epsilon,\tau )$), which is a generalization of the
Kolmogorov--Sinai entropy with finite resolution $\epsilon$, and where the
time is discretized by using a time interval $\tau$.

If the Kolmogorov--Sinai entropy can be calculated exactly and is finite, then
we can assure that the time--series was generated by a deterministic law.
Usually the $(\epsilon, \tau)$--entropy displays different behaviors as the
resolution is varied. According to these different behaviors one can
distinguish deterministic and stochastic dynamics. We can even define a
certain range of scales for these phenomena.

For a time--series long enough, the entropy can show a saturation range. For
$\epsilon \rightarrow 0$, one observes the following behaviors:
$h(\epsilon)\approx const $ for a  deterministic system, whereas
$h(\epsilon)\sim - \ln\epsilon$ for a stochastic system.
In general, predictability can be considered as a fundamental way to
characterize complex dynamical systems\cite{Boffetta}. We have used a method developed
in the papers\cite{Sugihara,Wales,Tsonis}, in order to investigate numerically
the randomness of
functions (4). This technique is very powerful  in distinguishing chaos
from random time series. The idea of the method is the following.
One can make short--term predictions that are based on a library
of past  patterns in a time series. By comparing the predicted and actual
values, one can make distinctions between random sequences and
deterministic chaos. For chaotic (but correlated) time series, the accuracy
of the nonlinear forecast falls off with increasing prediction--time interval.
On the other hand, for truly random sequences, the forecasting accuracy is
independent of the prediction interval. If the sequence values are correlated,
their future values may approximately be predicted from the behavior of
past values that are similar to those of the present. For uncorrelated random
sequences the error remains constant. The prediction accuracy is measured by
the coefficient of correlation between predicted and observed values. For
deterministic chaotic sequences this coefficient falls as predictions extend
into the future. Suppose we have a sequence $u_1,u_2,...,u_N$. Now we
construct a map with the dependence of $u_n(predicted)$ as a ``function'' of
$u_n(observed)$. If we have a deterministic chaotic sequence, this dependence
is almost a straight line, i.e. $u_n (predicted)\approx u_n(observed) $
(when the forecasting method is applied for one time step into the future).
When we increase the number of time steps into the future, this relation
becomes worse. The decrease with time of the correlation coefficient between
predicted and actual values has been used to calculate the largest Lyapunov
exponent of a time series\cite{Wales}. We have applied this method of
investigation to our functions(4)\cite{Jorge1}. When $Z$ is an integer $(Z>0)$, the
method shows that the function (4) behaves as a deterministic chaotic system.
If $Z$ is irrational, the correlation coefficient is independent of the
prediction time.
Even when the method is applied with prediction time interval $m=1$, the
correlation coefficient is zero (the map $(u_n(predicted),u_n(observed))$
covers completely the square $0\leq x \leq1$, $0\leq y \leq1$, showing
no patterns). This shows that the corresponding time series behaves as a
random sequence. When $Z=p/q$, function (4) behaves as a system with both,
deterministic chaos and noise. In this case, it is better to complement the
study with several alternative methods.

As pointed out by Cencini {\it et al.,}\cite{Cencini}, all these methods have in
common that one has to choose certain length scale $\epsilon$ and a particular
embedding dimension $m$. Thus the scenarios discussed in
\cite{Boffetta,Cencini}  can be very
useful in all the investigations aimed at the distinction between chaos and
noise.

\section{Conclusion}

Cohen\cite{Cohen} has reported that the solutions of chaotic ecological models
have  power spectra with increasing amplitudes at higher frequencies. This is
in contrast with the spectra presented in natural populations which are
dominated by low--frequency fluctuations.
Some authors\cite{Sugihara1}
suggest that this is a manifestation of  the interaction between biotic
factors and climatic factors. This problem  shows the difficulties in deciding
whether natural populations fluctuations  are determined by internal biological
mechanisms or they are mostly the  result of external environmental forcing.
We think that our results can help  to shed light on this issue. Recently
there have been reports\cite{Leirs,Dixon} on population dynamics where the
variability of the population originates  from both deterministic chaos and
stochastic processes. The complexity given  by equation (\ref{eq:2}) can help
to determine the relative weight of both factors.  In fact, the understanding
of the interaction of both deterministic and  stochastic processes is crucial
to model correctly the dynamics of an  ecological system.

We propose a combined approach to this issue: the SETAR model and the new
method for calculating complexity. In the particular case of the sheep
populations in the St. Kilda archipelago, it seems that the population
fluctuations are influenced mostly by frequent environmental variations
which include monthly wind, rain, temperature, food shortage and parisitism.

We believe that the ideas and methods used in the present article
can be
applied to other nonlinear random systems in biology and physics.




\begin{thebibliography}{}

\bibitem{Castro}
A. Castro-e-Silva and A. T. Bernardes,
Physica A 301 (2001) 63.

\bibitem{Dimitrova}
Z. I. Dimitrova and N. A. Vitanov,
Physica A 300 (2001) 91.

\bibitem{Droz}
M. Droz and A. Pekalski,
Physica A 298 (2001) 545.

\bibitem{Penna}
T. J. P. Penna, A. Racco and A. O. Sousa,
Physica A 295 (2001) 31.

\bibitem{Sznajd}
K. Sznajd-Weron and A. Pekalski,
Physica A 294 (2001) 424.

\bibitem{Sole}
R. V. Sol\'e, D. Alonso and A. McKane,
Physica A 286 (2000) 337.

\bibitem{Monetti}
R. Monetti, A. Rozenfeld and E. Albano,
Physica A 283 (2000) 52.

\bibitem{Johnson}
N. E. Johnson, D. J. T. Leonard, P. M. Hui and T. S. Lo,
Physica A 283 (2000) 568.

\bibitem{May1}
R. May,
Science 186 (1974) 645.

\bibitem{May2}
R. M. May,
{\it Stability and Complexity in Model Ecosystems} (Princenton  University Press,  NJ  1973).

\bibitem{Sugihara1}
G. Sugihara,
Nature (London) 378 (1995) 559.

\bibitem{Ellner}
S. Ellner, P. Turchin,  Am. Nat. 145 (1995) 343.

\bibitem{Sugihara2}
G. Sugihara,
Nature (London) 381 (1996) 199.

\bibitem{Grenfell1}
B. T. Grenfell {\it et al}.,
Nature (London) 394 (1998) 674.

\bibitem{Stenseth}
N. C. Stenseth and K. S. Chan,
Nature (London) 394 (1998) 620.

\bibitem{Paladin}
G. Paladin, M. Serva, and A. Vulpiani,
Phys. Rev. Lett.  74 (1995) 66.

\bibitem{Loreto}
V. Loreto, G. Paladin, and A. Vulpiani,
Phys. Rev. E 53 (1996) 2087.

\bibitem{Ranta}
E. Ranta, V. Kaitala, J. Lindst"m, and E. Helle,
Oikos 78 (1997) 136.

\bibitem{Grenfell2}
B. T. Grenfell, O. F. Price, S. D. Albon, and T.H. Clutton-Brock,
Nature (London) 355 (1992) 823.

\bibitem{Jorge1}
J. A. Gonz\'alez, L. I. Reyes and L. E. Guerrero,
Chaos 11 (2001) 1.

\bibitem{Jorge2}
J. A. Gonz\'alez, M. Mart\'{\i}n--Landrove, and L. Trujillo,
Int. J. Bifurcation and Chaos 10 (2000) 1867.

\bibitem{Jorge3}
J. A. Gonz\'alez and R. Pino,
Physica A 276 (2000) 425.

\bibitem{Jorge4}
H. N. Nazareno, J. A. Gonz\'alez, and I. F.  Costa,
Phys. Rev. B 57 (1998) 13583.

\bibitem{Wolf}
A. Wolf, J. B. Swift, H. L. Swinney, and J. Vastano,
Physica D 16 (1985) 285.

\bibitem{Pincus}
S. Pincus and B. S. Singer,
Proc. Natl. Acad. Sci. USA 93 (1996) 2083.

\bibitem{Pincus2}
S. Pincus and B. H. Singer,
Proc. Natl. Acad. Sci. USA 95 (1998) 10367.

\bibitem{Singer}
B. H. Singer  and S. Pincus,
Proc. Natl. Acad. Sci. USA 95 (1998) 1363.

\bibitem{Pincus3}
S. Pincus and R. E. Kalman,
Proc. Natl. Acad. Sci. USA 94 (1997) 3513.


\bibitem{Tong1}
H. Tong and K. S. Lim,
J. R. Stat. Soc. B 42 (1980) 245.

\bibitem{Tong2}
H. Tong,
{\it Non-linear Time Series: a Dynamical Systems Approach} (Oxford Univ. Press, 1990).

\bibitem{Tsay}
R. S. Tsay,
J. Am. Stat. Assoc. 84 (1989) 230.

\bibitem{Maynard}
S. J. Maynard, M. Slatkin,
Ecology 54 (1992) 384.



\bibitem{Sugihara}
G. Sugihara and R. M. May, Nature (London) 344 (1990) 734.

\bibitem{Wales}
D. J. Wales, Nature (London) 350 (1991) 485.

\bibitem{Tsonis}
A. A. Tsonis and J. B. Elsner, Nature (London) 358 (1992) 217.

\bibitem{Boffetta}
G. Boffetta, M. Cencini, M. Falcioni,and A. Vulpiani,
Physics Reports 356 (2002) 367.

\bibitem{Cencini}
M. Cencini, M. Falcioni, E. Olbrich, H. Kantz and A. Vulpiani,
Phys. Rev. E 62 (2000) 427.

\bibitem{Cohen}
J. E. Cohen,
Nature (London) 378 (1995) 610.

\bibitem{Leirs}
H. Leirs, {\it et al.},
Nature (London) 389 (1997) 176.

\bibitem{Dixon}
P. A. Dixon, M. J. Milicich, and G. Sugihara,
Science 283 (1999) 1528.

\end{thebibliography}
\end{document}